\documentclass[]{aa}
\usepackage{natbib} 
\usepackage{amssymb}
\usepackage{amsmath}
\usepackage{graphicx} 
\usepackage{txfonts} 
\begin{document} 
%
%
\title{
Resolving the compact dusty discs around binary post-AGB stars using
N-band interferometry \thanks{Based on observations made with the Very
Large Telescope Interferometer of the European Southern Observatory
(program id 073.A-9002(A)), the 1.2\,m Flemish Mercator telescope at
Roque de los Muchachos, Spain and the 1.2\,m Swiss Euler telescope at
La Silla, Chile } }
%
%
\author{
P. Deroo\inst{1} \and H. Van Winckel\inst{1} \and M. Min\inst{2} \and
L.B.F.M. Waters\inst{1,2} \and T. Verhoelst\inst{1} \and
W. Jaffe\inst{3} \and S. Morel\inst{4} \and F.  Paresce\inst{4} \and
A. Richichi\inst{4} \and P. Stee\inst{5} \and M. Wittkowski\inst{4} }

\institute{Instituut voor Sterrenkunde, K.U. Leuven, Celestijnenlaan
200B, B-3001 Leuven, Belgium \and Astronomical Institute ``Anton
Pannekoek'', University of Amsterdam, Kruislaan 403, 1098 SJ
Amsterdam, the Netherlands \and Leiden Observatory, P.B. 9513, Leiden
2300 RA, the Netherlands \and European Southern Observatory,
Karl-Scharzschild-Strasse 2, 85748 Garching, Germany \and Observatoire
de la C\^ote d' Azur, CNRS-UMR 6203, Avenue Copernic, 06130 Grasse,
France}
%
%
\date{Received <date> / Accepted <date>} 
\offprints{
P. Deroo \\ 
\email{Pieter.Deroo@ster.kuleuven.be}
} 
%
%
\abstract{
We present the first mid-IR long baseline interferometric observations
of the circumstellar matter around binary post-AGB stars. Two objects,
\object{SX Cen} and \object{HD 52961}, were observed using the
VLTI/MIDI instrument during Science Demonstration Time. Both objects
are known binaries for which a stable circumbinary disc is proposed to
explain the SED characteristics. This is corroborated by our N-band
spectrum showing a crystallinity fraction of more than 50\,\% for both
objects, pointing to a stable environment where dust processing can
occur. Surprisingly, the dust surrounding SX\,Cen is not resolved in
the interferometric observations providing an upper limit of 11 mas
(or 18 AU at the distance of this object) on the diameter of the dust
emission. This confirms the very compact nature of its circumstellar
environment. The dust emission around \object{HD\,52961} originates from a very
small but resolved region, estimated to be $\sim$35 mas at 8\,$\mu$m
and $\sim$55 mas at 13\,$\mu$m. These results confirm the disc
interpretation of the SED of both stars. In \object{HD\,52961}, the dust is not
homogeneous in its chemical composition: the crystallinity is clearly
concentrated in the hotter inner region.  Whether this is a result of
the formation process of the disc, or due to annealing during the long
storage time in the disc is not clear.

}
%
%
\titlerunning{
Discs around evolved objects: \object{HD\,52961} and SX\,Cen
}
\authorrunning{P. Deroo et al.} 
\maketitle

%
%
\section{Introduction}
The fast stellar evolution connecting the Asymptotic Giant Branch
(AGB) to the Planetary Nebulae (PNe) phase is still poorly understood
\citep[e.g.][]{Vanwinckel_2003}.  Many detailed studies of individual
transition objects (post-AGB stars) exist, but it is not clear how
these objects are related by evolutionary channels.  Moreover, there
is general agreement that binary interactions must play a significant
role in many well studied sources. Binarity is for instance invoked in the
physical models to understand the observational characteristics of
some spectacular geometries observed in PNe. More
recently, also the geometries and kinematical structures around
resolved post-AGB stars might be linked to binarity \citep[][ and
references therein]{Balick_2002}.  Since many uncertainties remain in
our understanding of the final evolution of single stars, it is no
surprise that this is even more the case when the
star is a member of a binary system.

Direct detection of the binary nature of central stars of resolved
nebulae is often difficult due to the high obscuration. Moreover, in
crossing the HR-diagram, the stars must pass the pop\,II Cepheid
instability strip, in which pulsational instabilities occur. This
makes radial velocity variations not a straightforward signature of
orbital variations.

In the sample of optically bright post-AGB stars, binaries are being
detected, however, and for an overview we refer to \citet[][ and
references therein]{Vanwinckel_2003}. One of the important
observational characteristics of those binaries is the shape of their
SED. They show a dust-excess starting near sublimation temperature,
irrespective of the effective temperature of the central object and
this despite the lack of a current dusty mass loss \citep[][ and
references therein]{Deruyter_2005, Deruyter_2006}.  Moreover, when
available, the long wavelength fluxes show a black-body slope which
indicates the presence of a component of large mm-sized grains. It is
argued in \cite{Deruyter_2006} that in all the investigated objects,
gravitationally bound dust is present, likely in a Keplerian disc.
Note that only for the most famous example, the \object{Red Rectangle}, this
dust emission is resolved and it shows a clear disc structure, both in
the near-IR and in the visible \citep{Menshchikov_2002,Cohen_2004}.
Moreover, the gaseous component was spatially resolved using
mm-interferometry \citep{Bujarrabal_2003,Bujarrabal_2005}. The latter
observations clearly demonstrated the Keplerian rotation of the
disc. In all other cases, the presence of a disc is postulated.  More
detailed studies of individual cases can be found: examples are
\object{89\,Her} \citep{Waters_1993}, \object{HR\,4049}
\citep{Waelkens_1991a, Dominik_2003} and \object{IRAS08544-4431}
\citep{Maas_2003}.  Given the orbits detected so far, one of the
conclusions is that it is clear that most binaries cannot have evolved
along single star evolutionary tracks.

The high spatial resolution of the mid-IR instrument MIDI mounted on
the VLTI interferometer of ESO makes this an ideal instrument to probe
the circumstellar material around these binaries for two reasons: (i)
the discs are likely compact so high spatial resolution measurements
are needed to resolve the discs and (ii) the discs are shown to emit a
significant part of their total luminosity in the N-band. We therefore
carefully selected 2 binary post-AGB stars for which there is
significant indirect evidence for the presence of a stable
circumstellar dust reservoir. The data presented in this contribution
are taken during Science Demonstration Time to illustrate the
potential of MIDI coupled to the VLTI to study the compact
circumstellar environment suspected in those evolved stars.

In Sect.~\ref{sect:global_characteristics}, we introduce both objects
and refine the orbital parameters published in our previous
papers. The observational log is presented in
Sect.~\ref{sect:observations} while the reduction of the
interferometric dispersed fringes is reported in
Sect.~\ref{sect:reduction}. We discuss our findings in 
Sect.~\ref{sect:discussion} and come to our conclusions in
Sect.~\ref{sect:conclusions}.


\section{\object{HD\,52961} and \object{SX\,Cen}: global
  characteristics}\label{sect:global_characteristics}

\object{SX\,Cen} and \object{HD\,52961}, having both a spectral type F-G
\citep{Kholopov_1999,Shenton_1994,Waelkens_1991}, are located in the
pop\,II instability strip with \object{SX\,Cen} known as a very regular
RV\,Tauri star with a period of 32.9 days, while \object{HD\,52961} shows a
photometric periodicity of 72 days. They are members of the chemically
anomalous post-AGB stars for which the photospheric abundances of the
different elements are closely linked to their condensation
temperature \citep{Waelkens_1991, Vanwinckel_1992, Vanwinckel_1995,
Maas_2002}.  Members of this class show higher photospheric abundances
for chemical elements with a lower condensation temperature.  In fact,
\object{HD\,52961} is one of the most extreme examples of this class of
objects. It is a highly metal-poor object \citep[\hbox{[Fe/H] = -4.8,
}][]{Waelkens_1991} which has more zinc than iron in absolute (!)
number \citep[\hbox{[Zn/Fe] = +3.1, }][]{Vanwinckel_1992}. There is
general agreement that this abundance pattern is caused by a chemical
fractionation process caused by dust formation in the circumstellar
environment. After decoupling of the gas and dust, reaccretion of the
gas causes the observed abundance pattern. \cite{Waters_1992} proposed
a scenario in which the circumstellar dust is trapped in a stable
disc. The occurrence of such a disc likely implies binarity for post-AGB
stars. Indeed, radial velocity measurements proved that all the
extremely depleted objects are binaries \citep{Vanwinckel_1995}. In
the following we refine our previously published Spectral Energy
Distribution (SED) as well as the orbital elements of both objects.

\subsection{Spectral energy distribution}\label{sect:SED}

\begin{figure}[t] \includegraphics[width=\hsize]{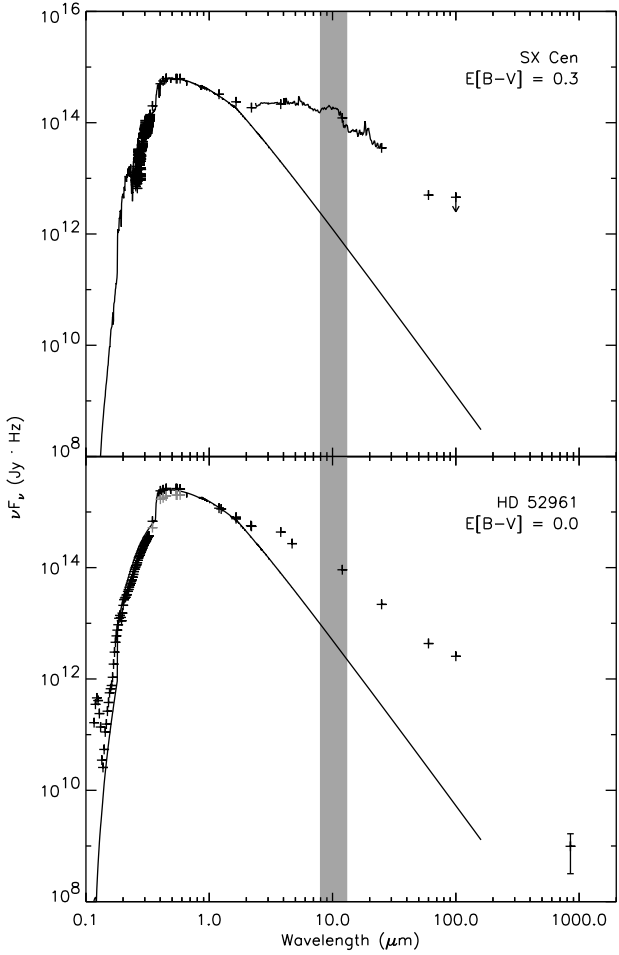}
\caption{The SED of both programme stars. On top, the SED of \object{SX\,Cen}
  is provided according to the parametrization given in the literature
  while at the bottom, the SED of \object{HD\,52961} is shown. The solid line is
  the scaled Kurucz model with stellar parameters determined in the
  literature (T$_{\rm{eff}} = 6500$\,K, log(g)$=1.5$ and [Fe/H]$=-1.0$
  for \object{SX\,Cen} \citep{Maas_2002} and T$_{\rm{eff}} = 6000$\,K,
  log(g)$=0.5$ and [Fe/H]$=-4.5$ for \object{HD\,52961}
  \citep{Waelkens_1991}). The photometric measurements are shown with
  plus symbols, black for photometric maximum, grey for photometric
  minimum. For \object{SX\,Cen}, the ISO/SWS spectrum is overplotted up to 25
  $\mu$m. The MIDI wavelength range is indicated by a grey box. }
\label{fig:sedBOTH} 
\end{figure}

The SED of \object{SX\,Cen} is discussed in the literature
\citep{Goldsmith_1987,Shenton_1994,Maas_2002} where a total reddening
of \hbox{E(B-V) = 0.3\,$\pm$\,0.1} is found. They find a broad
infrared excess starting already at K which, in combination with the
confirmed binarity, is interpreted in \cite{Maas_2002} as a signature
of a dusty disc, not an outflow. The SED is reproduced in the top
panel of Fig. \ref{fig:sedBOTH} in which the ISO/SWS spectrum is
overplotted. This spectrum shows a broad silicate emission feature at
10 $\mu$m, inside the MIDI wavelength range. Due to the low flux
levels at longer wavelengths, we cannot be conclusive about the origin
of the feature around 18 $\mu$m. This is possibly an artifact,
although it is present in both scans. The distance of \object{SX\,Cen} can be
estimated comparing the intrinsic luminosity with the integrated flux
of the scaled Kurucz model. The luminosity of \object{SX\,Cen} is estimated
using the period-luminosity relation derived for the LMC RV\,Tauri
variables \citep{Alcock_1998} to be about
\hbox{600\,$\pm$\,400\,L$_{\sun}$}. The large uncertainty is a relic
of the scatter in the P-L relation.  This provides a rough distance
estimate for \object{SX\,Cen} of \hbox{1.6\,$\pm$\,0.5\,kpc}.


For \object{HD\,52961}, we constructed a SED using IUE data (0.115 $\mu$m --
0.320 $\mu$m), Geneva optical photometry, near-IR JHKLM photometry
\citep{Bogaert_1994} and far-IR IRAS photometry. In addition, one SCUBA
\citep{Holland_1999} observation was made to obtain a continuum
measurement at 850 $\mu$m, providing F$_{850\mu\rm{m}} = 2.8 \pm
1.9$\,mJy with Mars as flux calibrator.

As the object suffers photometric variations over time, we constructed
the SED for photometric maximum only. The colour excess due to
interstellar and circumstellar extinction was estimated by searching
the best correspondence between the appropriate Kurucz model and the
dereddened SED in the optical and UV. The SED was dereddened using the
average interstellar extinction law of \cite{Savage_1979} and the
Kurucz model was chosen according to the stellar parameters given in
\cite{Waelkens_1991}, i.e. T$_{\rm{eff}} = 6000$\,K, log(g)$=0.5$ and
[Fe/H]$=-4.5$. The result is shown in Fig.~\ref{fig:sedBOTH}, where
the Geneva photometry at photometric minimum is overplotted using grey
crosses. While the photometry and the Kurucz model are consistent in
the UV and optical, a clear IR excess due to dust is observed at
longer wavelengths. This excess distribution, in combination with the
confirmed binarity \citep[][ and refinements in
Sect. \ref{sect:orbital_elements}]{Vanwinckel_1995} and the lack of a
current dusty mass loss, is also interpreted as evidence for a dusty
disc instead of an outflow. The luminosity of \object{HD\,52961} is estimated
as \hbox{1900\,$\pm$\,1300\,L$_{\sun}$} using the same
period-luminisoty relation as for \object{SX\,Cen}, providing a distance of
\hbox{1.4\,$\pm$\,0.5\,kpc}.

In the MIDI wavelength range \hbox{(8 -- 13 $\mu$m)}, the amount of
flux emitted by the stellar photosphere with respect to the total flux
is only 1\,\% for \object{SX\,Cen} and 5\,\% for \object{HD\,52961}. In addition, both
objects show a clear silicate resonance in emission in the N-band (see the
ISO/SWS spectrum in the top panel of Fig. \ref{fig:sedBOTH} and the
MIDI spectra in Fig. \ref{fig:totalspectrum}). Therefore, the MIDI
instrument, providing spectrally dispersed visibilities over the
N-band, is ideally suited to probe the circumstellar geometries of the
dust around both objects.

\subsection{Orbital elements}\label{sect:orbital_elements}
\begin{figure}[t]
\includegraphics[width=\hsize]{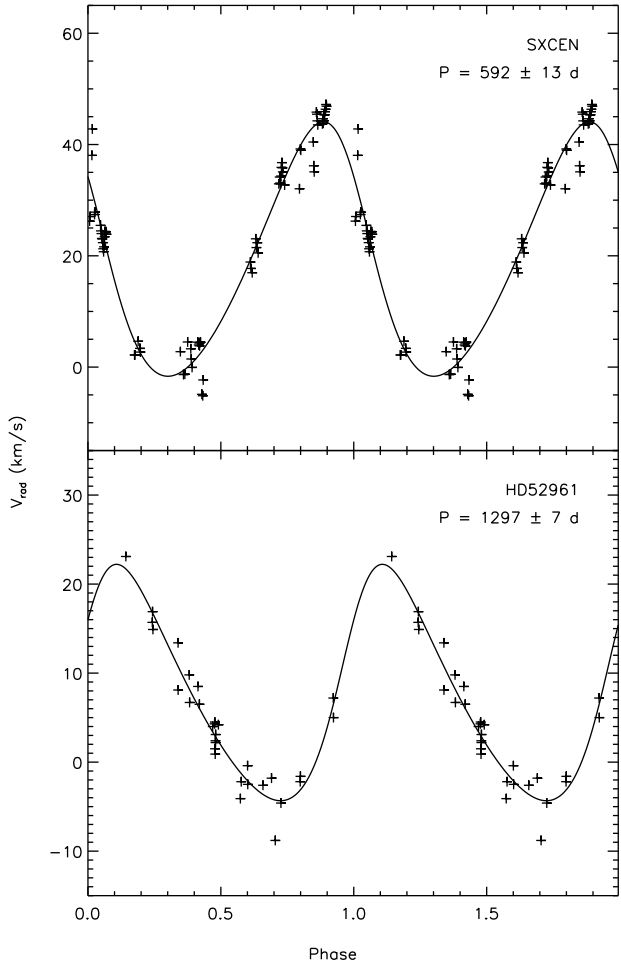}
\caption{The radial velocity curves of \object{SX\,Cen} and \object{HD\,52961} folded on
their binary period. The data of \object{SX\,Cen} were cleaned from
pulsationally induced variations as discussed in the text. The crosses
are the measurements while the solid line is the orbital solution.}
\label{fig:orbits}
\end{figure}

We refined the orbital elements which were published already in
\cite{Vanwinckel_1999} and \cite{Maas_2002} for \object{HD\,52961} and \object{SX\,Cen}
respectively. Our accumulation of data is now such that we covered
close to 3 (\object{HD\,52961}) and 2 (\object{SX\,Cen}) orbital cycles. The
heliocentric radial velocity data folded on the orbital periods are
given in Fig.~\ref{fig:orbits} and the orbital elements are listed in
Table~\ref{tab:orbitalelements}.

\begin{figure}[t]
\includegraphics[width=\hsize]{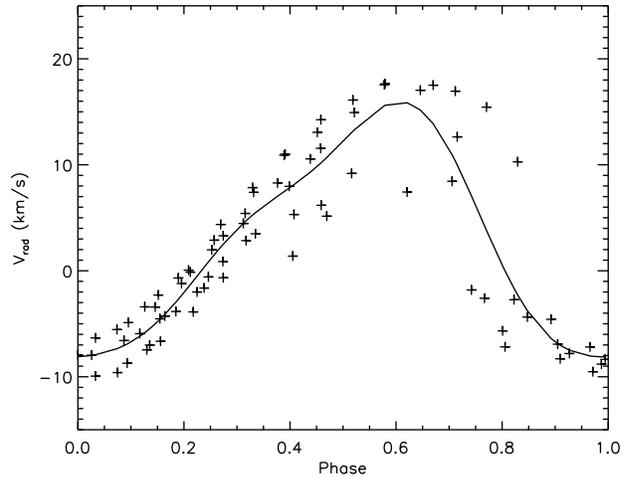}
\caption{The radial velocity variations induced by pulsation for
  \object{SX\,Cen}. The data indicated with crosses is pre-whitened with the
  orbital solution. The solid line is a harmonic least square fit
  of the 16.46 days pulsational period and three harmonics. }
\label{fig:pulsationSXCEN}
\end{figure}

The data sampling of \object{HD\,52961} is not very extensive and in the
residuals no clear modulation on the pulsational period is found.
For the RV\,Tauri star \object{SX\,Cen}, the pulsational amplitude
in radial velocity is significant. After pre-whitening of the orbital
solution, the pulsational period of 16.46\,d is clearly recovered
(Fig.~\ref{fig:pulsationSXCEN}).  We cleaned the original data with a
harmonic least square fit of the 16.46 days pulsation period and three
harmonics. The variance reduction of the pulsation model is 81\%.
After cleaning the original data with this pulsation model, we
redetermined the orbital elements. The eccentric orbit was found to be
significant according to the classical Lucy and Sweeney test
\citep{Lucy_1971}.

\begin{table}
\caption{The orbital elements of the program stars. All symbols have
their usual meaning. The number of measurements (N) and the
number of covered orbital cycles in our monitoring program are
also given.}\label{tab:orbitalelements}
\begin{tabular}{llll}
\hline \hline
                 & unit          &    \object{HD\,52961}       & \object{SX\,Cen} \\
\hline
P                & days          & 1297 $\pm$ 7       & 592 $\pm$ 13     \\
T$_{o}$          & JD            & 2448591 $\pm$ 38   & 2452107 $\pm$ 10 \\
K                & km\,s$^{-1}$  & 13.3 $\pm$ 0.9     & 22.9 $\pm$ 0.5   \\
$\gamma$         & km\,s$^{-1}$  & 7.4 $\pm$ 0.5      & 19.1 $\pm$ 0.4   \\
e                &               & 0.22 $\pm$ 0.05    & 0.16 $\pm$ 0.02  \\
$a\,\sin i$   & AU            & 1.54               & 1.23             \\
f(M)             & M$_{\odot}$   & 0.29               & 0.70             \\
N                & \#            & 31                 & 78               \\
cycles covered   &               & 2.9                & 2.0\\
\hline
\end{tabular}
\end{table}

As shown in our previous papers, both objects show a long term trend
in their photometric light-curve which is due to variable circumstellar
reddening in the line of sight towards the object.  For \object{SX\,Cen} this
long term trend is periodic with a period of 615 days
\citep{Oconnell_1933,Voute_1940}, very close to the orbital period.
For \object{HD\,52961} it was not very clear whether the subtle effect is
periodic or not.

If the circumstellar material is indeed mainly stored in a disc around
the objects and assuming this disc is located in the orbital plane,
the inclination of the disc cannot be very small. Assuming an
inclination varying between edge-on (i\,=\,90$^{o}$) up to 60$^{o}$,
and a mass of the evolved component of 0.6 M$_{\odot}$, the mass of
the companion varies between 0.8\,--\,1.1 M$_{\odot}$ for \object{HD\,52961}
and 1.4\,--\,1.9 M$_{\odot}$ for \object{SX\,Cen}. The unseen companion is
probably an unevolved main sequence star, since the lack of an UV
excess and of any sign of symbiotic activity make the presence of a
massive compact object very unlikely. Moreover, the orbital
characteristics in period and eccentricity make it very unlikely that
the companion is a post red giant as well \citep{Vanwinckel_2003}.

Neither stars are filling their Roche Lobe now, but in both cases it
is clear that the actual orbit is too small to accommodate a full
grown AGB star. The stars must have suffered an evolutionary phase
with severe binary interaction when at giant dimensions.

%
%
\section{Observations}\label{sect:observations}
The VLTI/MIDI interferometer \citep{Leinert_2003} was used to combine
the light coming from the UT2 and UT3 telescopes. The observations of
the targets, \object{SX\,Cen} and \object{HD\,52961}, were performed in three nights of
Science Demonstration Time in February at a projected baseline in the
range of 40 to 50 meters. A detailed log of the observations of the
science targets is presented in Table \ref{tab:log}.

The following observing sequence was carried out, according to the
standard procedures for MIDI, and repeated for target stars and
calibrators. First, acquisiton images are obtained by both telescopes
independently (i.e. without beam combiner and prism) to ensure overlap
of the beams, which is required for interferometric combination. Then,
the MIDI beam combiner, the slit and the prism are inserted. This
produces two spectrally dispersed interferometric outputs of opposite
phase. The zero optical path difference (OPD) is searched for by
scanning a range of a few millimeters around the expected value. When
found, MIDI uses its piezo-driven mirrors to keep the fringe pattern
at a fixed position within a $\approx 200$\,$\mu$m scan length, while the
VLTI delay lines compensate for the drift in OPD position due to
sidereal motion and for the slow component of atmospheric
piston. Fringes are integrated for about 1-3 minutes. Finally,
photometric data are recorded using one telescope at a time, with the
same optical set-up but using chopping to subtract sky and background.
%

To correct for optical imperfections and atmospheric turbulence, a
calibrator of known diameter is measured as well. The time-lag between
the measurement of this calibrator and the science object is about 30
min. However, considering the present accuracy per single visibility
measurement of about 10\,\%, we can also use calibrators observed in
the same mode one or two hours earlier or later \citep[see
e.g.][]{Leinert_2004}. A list of the calibrator observations is given
in Table \ref{tab:log}.

\begin{table*} 
\caption{A summary of the observations with the MIDI instrument of
 \object{SX\,Cen} and \object{HD\,52961}. For each science target, the calibrators used
 to calibrate the visibility are given (the flux calibrators are given
 in Table \ref{tab:photcal}).  The angular diameter in the Limb
 Darkened Disc approximation is obtained from \cite{Verhoelst_2005}
 (cf. http://www.ster.kuleuven.ac.be/$\sim$tijl/MIDI\_calibration/mcc.txt).
 The reported flux for the calibrator sources is the IRAS 12.5\,$\mu$m
 flux. Nomenclature: UT = Universal Time, PB = Projected Baseline and PA
 = Projected baseline Angle}
\label{tab:log}
\begin{center} 
\begin{tabular}{l c c c c c c c c c c} \hline \hline
\multicolumn{1}{c}{night} & 
\multicolumn{1}{c}{science} &
\multicolumn{1}{c}{UT} & 
\multicolumn{1}{c}{PB} &
\multicolumn{1}{c}{PA} &
\multicolumn{1}{c}{airmass} &
\multicolumn{1}{c}{calibrator} &
\multicolumn{1}{c}{UT}     &
\multicolumn{1}{c}{spectral} &
\multicolumn{1}{c}{diameter} &
\multicolumn{1}{c}{flux} \\

\multicolumn{1}{c}{yyyy/mm/dd} &
\multicolumn{1}{c}{target} & 
\multicolumn{1}{c}{hh mm ss} &
\multicolumn{1}{c}{(m)} & 
\multicolumn{1}{c}{(\degr )} &
\multicolumn{1}{c}{}&
\multicolumn{1}{c}{target} &
\multicolumn{1}{c}{hh mm ss}     &
\multicolumn{1}{c}{type} &	
\multicolumn{1}{c}{(mas)} &
\multicolumn{1}{c}{(Jy)} \\
\hline
2004/02/09 &
  \object{HD\,52961} & 02 20 02 & 39.7 & 45 & 1.2
     & HD\,49161  & 02 45 52 & K4\:III   & 2.44 $\pm$ 0.01 & 10.35\\ &
  \object{SX\,Cen}   & 07 37 52 & 44.6 & 41 & 1.1
     & HD\,67582  & 06 24 15 & K3\:III   & 2.30 $\pm$ 0.01 & 9.33 \\ & & & & &
     & HD\,67582  & 07 13 39 & K3\:III   & 2.30 $\pm$ 0.01 & 9.33 \\ & & & & &
     & HD\,107446 & 08 08 53 & K3.5\:III & 4.43 $\pm$ 0.02 & 32.42 \\
2004/02/10 &
  \object{HD\,52961} & 04 13 04 & 46.1 & 46 & 1.4
     &  HD\,67582  & 03 46 12 & K3\:III  & 2.30 $\pm$ 0.01 & 9.33 \\ &
  \object{SX\,Cen}   & 07 37 52 & 44.6 & 41 & 1.1
     & HD\,107446 & 08 16 44 & K3.5\:III & 4.43 $\pm$ 0.02 & 32.42 \\ 
2004/02/11 &
  \object{SX\,Cen}   & 08 03 25 & 43.6 & 46 & 1.1
     & HD\,120404 & 08 27 38 & K7\,III   & 2.96 $\pm$ 0.02 & 13.28 \\
\hline
\end{tabular}
\end{center}
\end{table*}

%
%
\section{Reduction}\label{sect:reduction}

\subsection{incoherent vs coherent analysis}
We used two different methods for the MIDI data reduction. The first
method is based on power spectrum analysis (hereafter called
incoherent analysis), while the second method reduces all frames to
the same OPD and adds them coherently (hereafter called coherent
analysis). For the incoherent analysis of the data, we used the MIA
package (MIDI Interactive Analysis,
http://www.mpia-hd.mpg.de/MIDISOFT/) developed at the Max-Planck
Institut f\"ur Astronomie in Heidelberg, while for the coherent
analysis we used the EWS package (Expert Work Station) developed by
Walter Jaffe at the Leiden observatory \citep{Jaffe_2004}.

During the incoherent analysis, we separated the different scans in
those with and without fringes, where each scan is Fourier-transformed
from OPD to fringe frequency space. Considering the wavelengths
present in the band and the rate at which the OPD is changing, the
power is calculated in the correct frequency interval. The total power
of all measured scans with fringes is then averaged and an estimate of
the noise is subtracted. This noise estimate is based on the frames
without fringes. This provides a value of the instrumental visibility
squared of each channel. Contrary to the coherent method, the major
difficulty of this method is that an accurate estimate of the
off-fringe noise power is needed. Since our science targets have small fluxes in
the 10 $\mu$m window, a reliable estimate of the noise power is
difficult to obtain and we focused during data reduction on the
coherent method (see below). Our incoherent analysis was only used to
check the results obtained by a coherent analysis and both methods
give consistent results.

\subsection{coherent analysis}
We first investigated the photometric datasets. The averages of the
target and sky frames are calculated and subtracted, providing a raw
two-dimensional spectrum of the object. The position and width of this
spectrum is determined and a spatial mask is constructed from the
location and average width of the spectrum at each wavelength
position. After multiplication of the detector images with this mask,
the rows are added providing a one dimensional raw spectrum of
the object (i.e. not corrected for the atmospheric transmission and
instrumental efficiency).

The spatial mask is then used to extract the information of the
interferometric observations as well, in the assumption that all
instrumental parameters stay the same between the interferometric and
photometric observation. The two detector spectra with opposite phase,
are subtracted, resulting in one interferometrically modulated
spectrum.  In this way, the background is reduced by approximately 90
percent.

Contrary to the incoherent method which allows the summing of scans
where the relative OPD is not known, the coherent method needs an
accurate determination of the atmospheric delay. The large wavelength
coverage of the N-band ensures that this can be accurately done by
measuring the fringes in frequency space \citep[rather than in OPD
space which is done in an incoherent analysis, e.g.][]{Tubbs_2004}. As
a first step, the known instrumental delay is removed from each frame
after which the (previously unknown) atmospheric delay is retrieved
using a group delay estimation. At this point, the data is not yet
fully coherent because of the instrumental phase imposed on the data
(e.g. the varying index of refraction of water vapor imposes
variations in phase that are not removed by a group delay
fitting). These phase shifts are almost constant as a function of
frequency and can be approximated as a constant phase shift over the
N-band \citep{Jaffe_2004}. Finally, the data can be added coherently
to obtain the final visibility amplitude and differential phase.

%
%

%

The instrumental visibility is then calculated dividing the fringe
amplitude by the non-interferometric, photometric exposures.
Repeating this procedure for a calibrator enables to estimate the
instrumental visibility loss and thus determining the calibrated
visibility of the science object.


\subsection{the data}

\subsubsection{photometry}\label{sect_photometry}

\begin{figure}[t]
\includegraphics[width=\hsize]{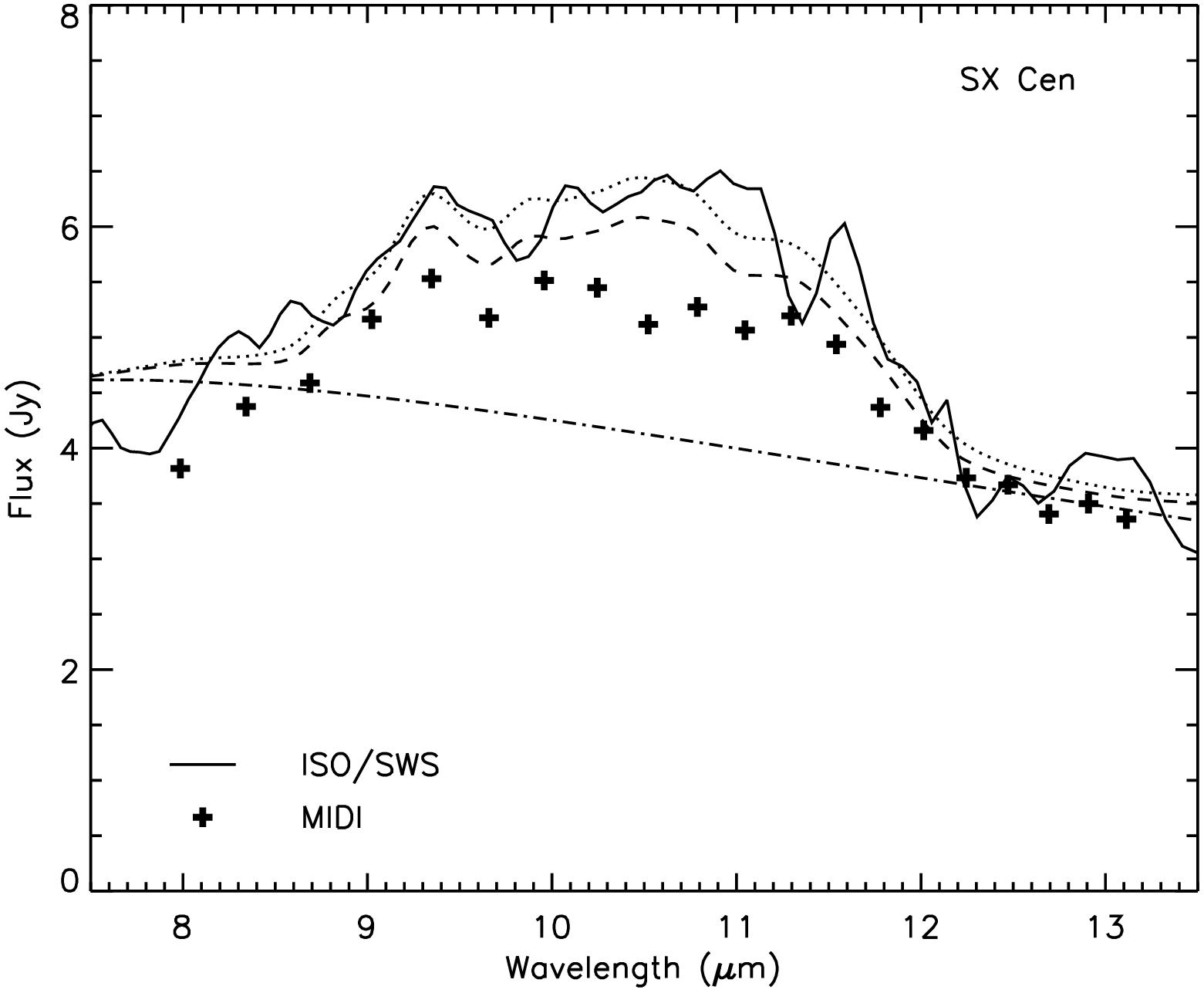}
\includegraphics[width=\hsize]{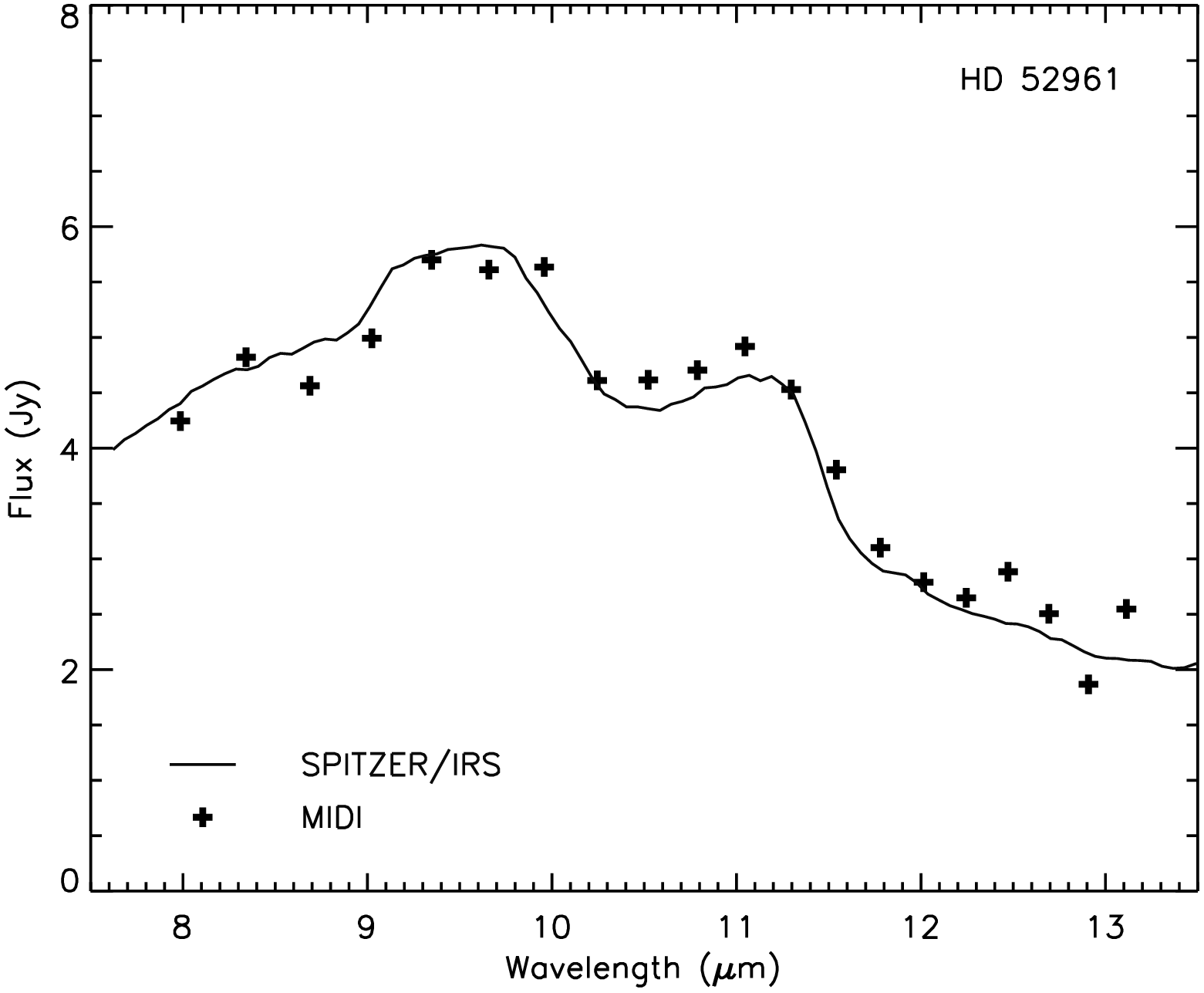}
\caption{The MIDI single telescope spectra of both programme
  stars are shown using black crosses. For the flux calibration, we
  only used the data of 10 February. In the upper panel (SX Cen), the
  full line shows the ISO/SWS spectrum. The dotted line is the result
  of our chemical model (Sect. \ref{sec:chemical_composition}), while
  the contribution of different individual dust components of the fit
  are given as well: the dashed-dotted line for the continuum and the
  dashed line for the crystalline component. In the lower panel, the
  full line shows the SPITZER/IRS spectrum of \object{HD\,52961}. }
\label{fig:totalspectrum}
\end{figure}

A raw spectrum is obtained each night by subtracting the masked target
frames from the masked sky frames. This spectrum is flux calibrated
and corrected for atmospheric transmission using the calibrator
spectra observed during the same night.  For the calibrators, the
intrinsic spectra were synthetised from {\sc marcs} atmosphere models
\citep[][ and further updates]{Gustafsson_1975}, using the
temperature, surface gravity and angular diameter determined in
\cite{Vanboekel_2004}. This approach is preferred over a
Rayleigh-Jeans approximation of the calibrator spectrum, since the SiO
first overtone band head is not negligible in a K giant N-band
spectrum. The 12\,$\mu$m flux of this synthetic spectrum is well
within 1\,$\sigma$ of the color corrected IRAS 12\,$\mu$m flux.  For
both objects, the absolute flux calibration has been performed with
the data of 10 February only, using the calibrators listed in Table
\ref{tab:photcal}. Both reduced spectra (R\,$\sim 30$) are shown in
Fig.~\ref{fig:totalspectrum}, where they are compared to independent
spectra taken by the ISO/SWS (R\,$\sim 248$) and the SPITZER/IRS
(R\,$\sim 127$) instrument.

\begin{table} 
\caption{A summary of the calibrators used to flux calibrate the
  spectrum of \object{SX\,Cen} and \object{HD\,52961}. All calibrators listed were
  observed on February, 10. The reported flux is the IRAS 12.5 $\mu$m
  flux, and the diameters are taken from \cite{Verhoelst_2005}. }
\label{tab:photcal}
\begin{center} \begin{tabular}{l c c c c c} \hline \hline
\multicolumn{1}{c}{calibrator} &
\multicolumn{1}{c}{UT} & 
\multicolumn{1}{c}{airmass} &
\multicolumn{1}{c}{spectral} &
\multicolumn{1}{c}{diameter} &
\multicolumn{1}{c}{flux} \\

\multicolumn{1}{c}{target} & 
\multicolumn{1}{c}{hh mm} &
\multicolumn{1}{c}{} &
\multicolumn{1}{c}{type} &	
\multicolumn{1}{c}{(mas)} &
\multicolumn{1}{c}{(Jy)} \\
\hline
HD\,67582 & 02 37 & 1.09 & K3 III & 2.30 $\pm$ 0.01 & 9.33 \\
HD\,67582 & 03 46 & 1.07 & K3 III & 2.30 $\pm$ 0.01 & 9.33 \\
HD\,49161 & 04 43 & 1.56 & K4 III & 2.44 $\pm$ 0.01 & 10.35 \\
HD\,107446 & 08 16 & 1.24 & K3.5 III & 4.43 $\pm$ 0.02 & 32.42 \\
\hline
\end{tabular}
\end{center}
\end{table}

\subsubsection{interferometry}\label{sec:interferometric_measurement}

We start this discussion with an error estimate on the observed
visibilities. The main source of error is the varying overlap between
the interferometric beams due to imperfect source acquisition and
residual image motion \citep[see e.g.][]{Leinert_2004}. This reduces
the visibility for calibrator and/or science object with an unknown
amount. The shape over the N-band, however, remains the same. The
visibility variation within the spectral band,
is therefore much more reliable than its absolute value.
 
\begin{figure}[t]
\includegraphics[width=\hsize]{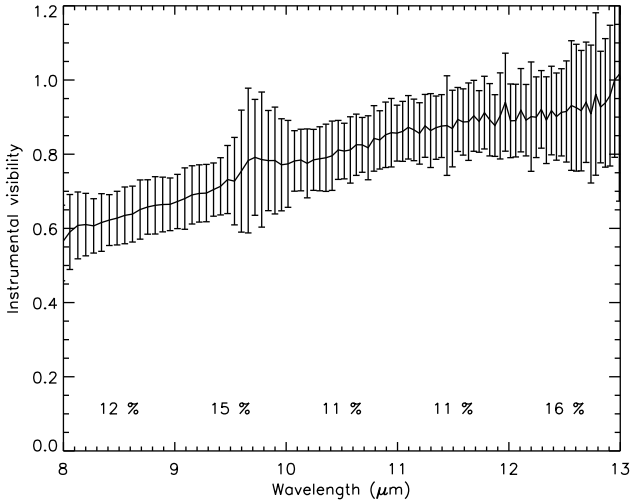}
\caption{The mean instrumental visibility of the six calibrators taken
  in prism mode on February 9. The variance on the mean is overplotted
  and for each wavelength interval of 1 $\mu$m, the mean variance is
  written below. The variance gives an estimate of the absolute error
  on the visibility, however the relative visibility is more reliable
  ($\sim$\,5\%). }
\label{fig:instrumental_visibility} \end{figure}

To get a quantitative estimate of the absolute uncertainty on the
visibility, we look at all calibrators ($\sim$\,point sources)
observed during one night. If the interferometric efficiency is
constant throughout the night, all calibrator measurements should
yield the same instrumental visibility. In
Fig. \ref{fig:instrumental_visibility}, the mean instrumental
visibility of all six calibrators observed in the prism mode during
the night of February 9 is plotted. The variance on the mean is
overplotted. It is clear from this figure that the instrumental loss
of visibility is much higher at 8 $\mu$m than at 13 $\mu$m and that
the uncertainty on the absolute value of the visibility is about 15
\%. However, when calibrating the visibility of the science source
using a calibrator source observed in direct concatenation, this
quantitative error is an upper limit. In the following, we use an
error of 15 \% on the absolute visibility, which is therefore a
conservative estimate.

Calibrated visibilities are obtained dividing the raw visibility by
the instrumental visibility. To calibrate the measurement of \object{SX\,Cen}
observed at 9 February, we used the mean instrumental visibility as
obtained from the three last calibrator measurements. Unfortunately,
such a mean could not be used for the other measurements. Instead, we
employed the calibrator closest in time to calibrate the visibility of
the science source (see Table \ref{tab:log}). The resulting calibrated
visibilities are shown in Figs.~\ref{fig:meanVisSXCEN} and
\ref{fig:visibilities_reduced}.

%
%
\section{Discussion}\label{sect:discussion}

\subsection{visibilities}
Because the angle between the projected baselines is the same to
within 5 degrees for both objects, no large effects due to a possible
asymmetry in the source morphology are expected. Therefore, as a
first-order approximation, we modelled the circumstellar environment
of objects using a uniform disc. The visibility in this assumption is
given by $V = 2 J_1(x)/x$, where $x=2\pi \theta B/ \lambda$ with
$\theta$ the diameter of the disc and $B$ the projected baseline
length. This function is smoothly increasing with wavelength, as
long as $x < 1.22 \pi$.
The increase is however different for various amounts in resolving
power, a steeper increase is observed if the source is more
resolved. Assuming a temperature distribution in the disc, with the
colder dust located further away from the star than the hotter dust,
the gradient decreases. For an unresolved source, the value of the
visibility remains constant at unity.



\begin{figure}[h]
\includegraphics[width=\hsize]{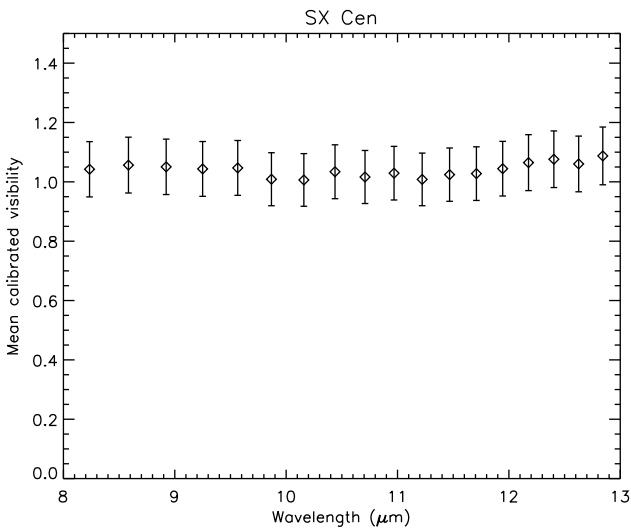} 
\caption{The mean calibrated visibility of \object{SX\,Cen}. The
  value close to unity and the flatness of the profile over the passband
  show that \object{SX\,Cen} is unresolved at the particular orientation and
  baseline setting. }
\label{fig:meanVisSXCEN} \end{figure}

The disc around \object{SX\,Cen} is unresolved in all measurements even using a
45\,m baseline. The visibility is close to unity and shows a flat
distribution over the passband. The mean calibrated visibility for all
measurements is shown in Fig. \ref{fig:meanVisSXCEN}. Because all
measurements are observed at approximately the same projected angle
(ranging from 41 to 46 degrees) this means that in that particular
orientation, the structure is smaller than 11 mas at 8 $\mu$m and 17
mas at 13 $\mu$m in a uniform disc approximation.  Using a gaussian
distribution modelling, the FWHM gives respectively 7 mas and 10 mas as
upper limits.

\begin{figure}[t] 
\includegraphics[width=\hsize]{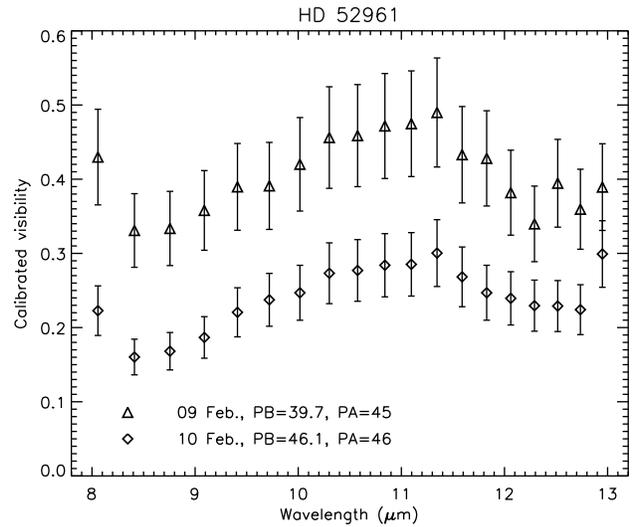} 
\caption{The calibrated visibilities for \object{HD\,52961} as obtained in
  both measurements.  The projected baseline (PB) is given in
  meters and the projected angle (PA) in degrees. Different symbols
  were used for both measurements.}
\label{fig:visibilities_reduced}
\end{figure}

\begin{figure}[h]
\includegraphics[width=\hsize]{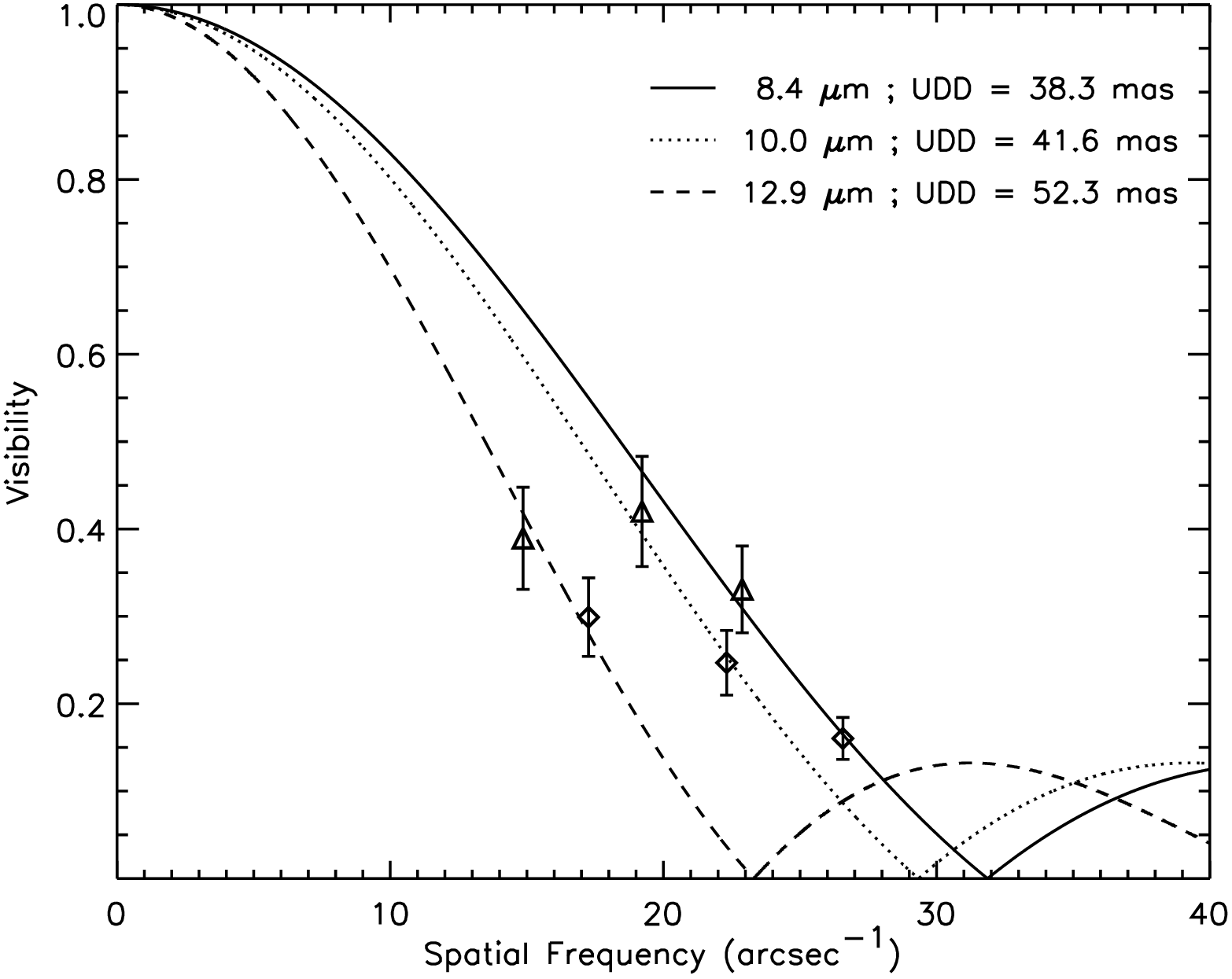}
\includegraphics[width=\hsize]{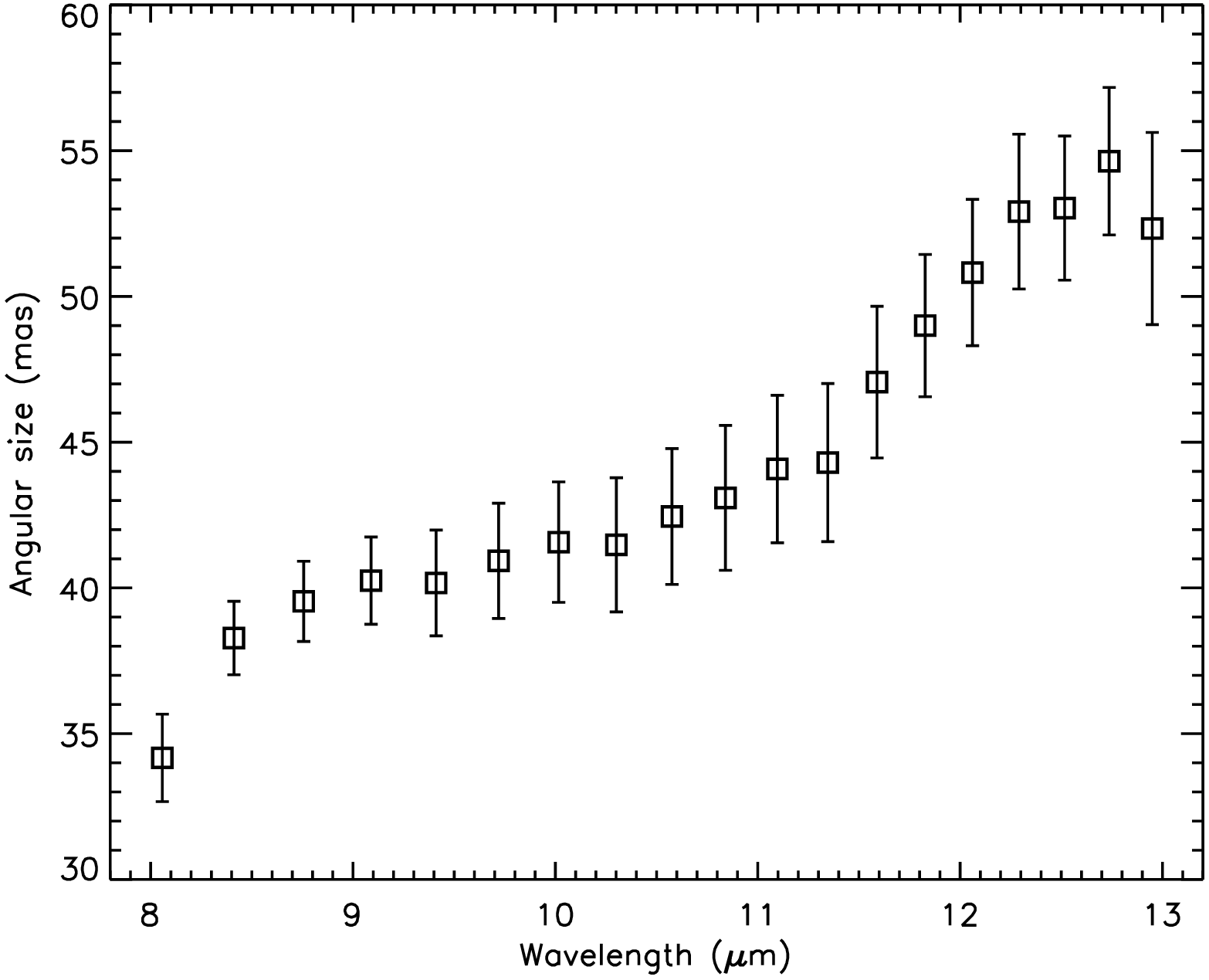} 
\caption{\emph{Upper panel:} The uniform disc fit to the data for
  three different wavelength bins. The lines in the plot are the
  uniform disc models with different sizes while the symbols are the
  measurements. Triangles were used for the measurements of 9 February
  while diamonds were used for those of 10 February. \emph{Lower
  panel:} The angular size of the dust around HD52961 as determined
  using a uniform disc model for each wavelength bin.}
\label{fig:angularSize} \end{figure}

\object{HD\,52961} shows quite a different picture. For this source, the
visibilities are low (see Fig.~\ref{fig:visibilities_reduced}), thus
the source is clearly resolved. Immediately noted is the fact that we
do not get an increase in visibility amplitude which is quite linear
(expected for a uniform disc model in the observed visibility
range). Instead we see a ``bump'' in the visibility pattern ranging
from 9 to 12\,$\mu$m. The geometry of the disc around this object can
clearly not be modelled with the same uniform disc at all wavelenghts
(see also Fig. \ref{fig:angularSize}). Using a uniform disc
approximation for each wavelength independently is however
instructive. For each wavelength bin, we made a $\chi^2$
minimalisation between the observed visibility at both baselines and a
uniform disc model. This fit is shown for three representative
wavelengths in the upper panel of Fig. \ref{fig:angularSize}. The
diameter of the source at all wavelength bins in a uniform disc
approximation is shown in the lower panel of
Fig. \ref{fig:angularSize}. The measurements at both baselines are
very consistent in a uniform disc approximation for each wavelength
independently (the mean reduced chi-square over the wavelength band is
as low as 0.09) and provide a diameter increasing from $\sim$35 mas at
8 $\mu$m to about $\sim$55 mas at 13 $\mu$m. In a gaussian
distribution modelling, the FWHM gives respectively $\sim$23 mas and
$\sim$34 mas (and a mean reduced chi-square of 0.22). The increase
towards longer wavelengths is consistent with a dust-distribution for
which the temparture decreases further away from the star. We however
note that the observed increase in size is not smooth over the
wavelength band. There is an increase in size from 8 $\mu$m to 8.5
$\mu$m and onwards 11.5 $\mu$m, while in between a sort of plateau
exists. We interpret this plateau as resulting from a non-homogenous
distribution of the radiating silicates which contribute most in the
inner regions close to the central star, thus lowering the overall
size (see also the following sections).

For both objects we interpret the small angular scales of the dust
around the objects as another clear indication that the circumstellar
dust is stored in a compact Keplerian disc around the system.
 
\subsection{spectra} 
The SED of both objects shows a significant near-IR excess, indicative
of a hot dust component, while there is no evidence for an ongoing
dusty mass loss of those rather hot stars. In \cite{Deruyter_2006},
this is interpreted as originating from a hot inner rim near dust
sublimation temperature. Because no dust can survive at higher
temperatures, this dust receives head on radiation from the star and
is therefore supposed to be puffed up \citep[e.g. the wall model for
HR4049 elaborated by][]{Dominik_2003}. This is corraborated by the
lower limit on the opening angle of the disc as seen from the star of
13$^o$ for \object{HD\,52961} and 32$^o$ for \object{SX\,Cen} \citep{Deruyter_2006}. In
this model, we expect a highly centerally peaked intensity
distribution which provides that the correlated spectra measured by
the interferometer are dominated by the inner regions of the
disc. However, because of the varying spatial resolution from 8 to 13
$\mu$m, a slope is introduced in the correlated spectrum. To determine
the magnitude of this effect, a detailed modelling has to be
performed, which is out of the scope of this article. For now, we
assume that this effect is a smooth function of wavelength, thus
having only a marginal effect on any mineralogy determination.



\begin{figure}[t]
 \includegraphics[width=\hsize]{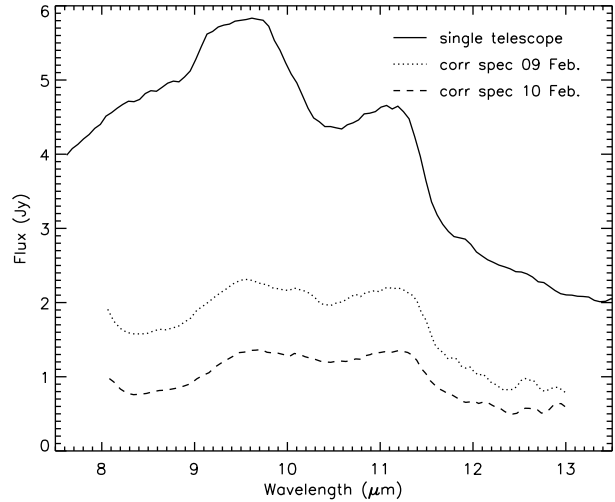}
 \caption{The spectra of different parts of the dust around
   \object{HD\,52961}. The solid line is the single telescope spectrum sampling all
   the dust around \object{HD\,52961}. The dotted and dashed line give the
   spectrum of smaller parts of the disc around the object, namely the
   unresolved part of the disc in each measurement.}
 \label{fig:allspectraatonce}
 \end{figure}

Because \object{SX\,Cen} is unresolved at all employed baseline settings, the
correlated spectrum is identical to the single telescope spectrum and
thus no additional information is available for this object. However,
for \object{HD\,52961}, which is clearly resolved in both measurements, the
shape of the two correlated spectra is predominantly determined by the
inner parts of the disc. This means that we can construct independent
spectra of geometrically different areas of dust around \object{HD\,52961}. The
single telescope spectrum provides the full N-band spectrum of all the
dust around \object{HD\,52961}. The correlated spectra sample smaller parts of
the disc. These spectra, each sampling a different geometrical part of
the disc, are shown in Fig. \ref{fig:allspectraatonce}. From this
figure, it is clear that the shape of the correlated spectra is quite
different from the single telescope spectrum. This points to a
different chemical composition of the inner part of the disc and the
outer part of the disc. To quantify this, we have fitted the different
spectra independantly.

\subsection{silicate mineralogy}\label{sec:chemical_composition}
In order to determine the mineralogy and sizes of the emitting dust
grains, we made a fit to the N-band spectra using calculated
emissivities of irregularly shaped, chemically homogeneous dust
grains. The most important dust species causing spectral signature in
the 10\,$\mu$m window are amorphous and crystalline olivine
(Mg$_{2x}$Fe$_{2-2x}$SiO$_4$), amorphous and crystalline pyroxene
(Mg$_{x}$Fe$_{1-x}$SiO$_3$), and amorphous silica (SiO$_2$), where $x$
determines the Mg/Fe ratio ($x=1$ for the crystalline silicates,
$x=0.5$ for the amorphous silicates). The complex refractive indices
for the different grain species were taken from various authors listed
in Table \ref{tab:refractive_indices}. To simulate the effects of
particle irregularity we employ a particular implementation of the
so-called \emph{statistical approach} using a distribution of hollow
spheres. This distribution is very successful in reproducing the
measured absorption spectra of irregularly shaped particles
\citep{Min_2003, Min_2005}. In addition to the dust species causing
the feature, we also add a continuum contribution which accounts for
emission by large grains and/or for the possible presence of
featureless components such as metallic iron and iron sulfide. This
continuum contribution is modeled using a constant mass absorption
coefficient. In the 10 $\mu$m region we are mainly sensitive to the
dust grains smaller than a few $\mu$m. We represent the size
distribution of the particles by two different grain sizes, $0.1$ and
1.5\,$\mu$m. A similar method was successfully employed by, for
example, \citet{Bouwman_2001}, \cite{Honda_2003},
\cite{Honda_2004}, \cite{Vanboekel_2004nature} and
\cite{Vanboekel_2005} to fit 10\,$\mu$m emission spectra of
circumstellar discs. Particles larger than a few $\mu$m contribute
mainly to the continuum. In addition we assume that the thermal
radiation we analyze originates from optically thin parts of the disc,
which allows us to add the contributions from the various components
linearly.  For the emission of the outer parts of the disc,
tentatively attributed to layers directly heated by the stellar flux,
this is a reasonable approximation. Because the stellar radiation is
incident under a high inclination, the temperature distribution in the
surface layer of the disc must be very sharp and therefore, the
emission in the N-band comes likely from optically thin parts. For the
inner parts of the disc, the situation is more complex: a large
fraction of the radiation comes from the inner rim, which has regions
of both low and high opacity. The fit, using an optically thin
assumption for the different contributing minerals, is therefore
certainly too simplistic.  We use it here as a first order estimate to
show the chemical gradient of the silicates in the disc, but a
detailed 2D radiative transfer model with a gradient in the
physico-chemical condition of the dust grains will be needed to
quantify the results. This is outside the scope of this paper.

We assume all dust grains, including the ones causing the continuum,
to have the same temperature distribution. Due to the limited
wavelength range this temperature distribution can be represented by a
single Planck curve with a characteristic temperature $T_c$. This is
justified because it is very likely that the dust grains of different
species are coagulated, implying thermal contact between the various
components. The characteristic temperatures used in the modeling are
given in Table \ref{tab:composition}.

\begin{table}[!t]
\begin{center}
\linespread{1.3}
\selectfont
\begin{tabular}{lc}
\hline
\hline
grain species & reference \\
\hline
Amorphous Olivine       & \cite{Dorschner1995} \\
Amorphous Pyroxene      & \cite{Dorschner1995} \\
Forsterite    & \cite{Servoin1973}   \\
Enstatite     & \cite{Jaeger1998}    \\
Amorphous Silica        & \cite{1960PhRv..121.1324S} \\
\hline
\end{tabular}
\end{center}
\linespread{1}
\caption{A list of the references of the complex refractive indices employed
  for the various grain species. }
\label{tab:refractive_indices}
\end{table}

The abundances of the dust components are determined by using a linear
least square fitting procedure with constraints on the weights to
avoid negative values. The temperature of the grains and the
underlying continuum is varied from 0 to 1500\,K until a best fit is
obtained.

The dust parameters derived from the unresolved spectrum of \object{SX\,Cen}
are given in the upper row of Table~\ref{tab:composition}. 
The resulting best fit spectrum is shown as a dotted line in
Fig.~\ref{fig:totalspectrum}. The grains in the circumstellar
environment of \object{SX\,Cen} are highly crystalline and also on average
relatively large compared to the interstellar grain population. This
implies a large amount of dust processing in the circumstellar
environment.

\begin{figure}[t]
\includegraphics[width = \hsize]{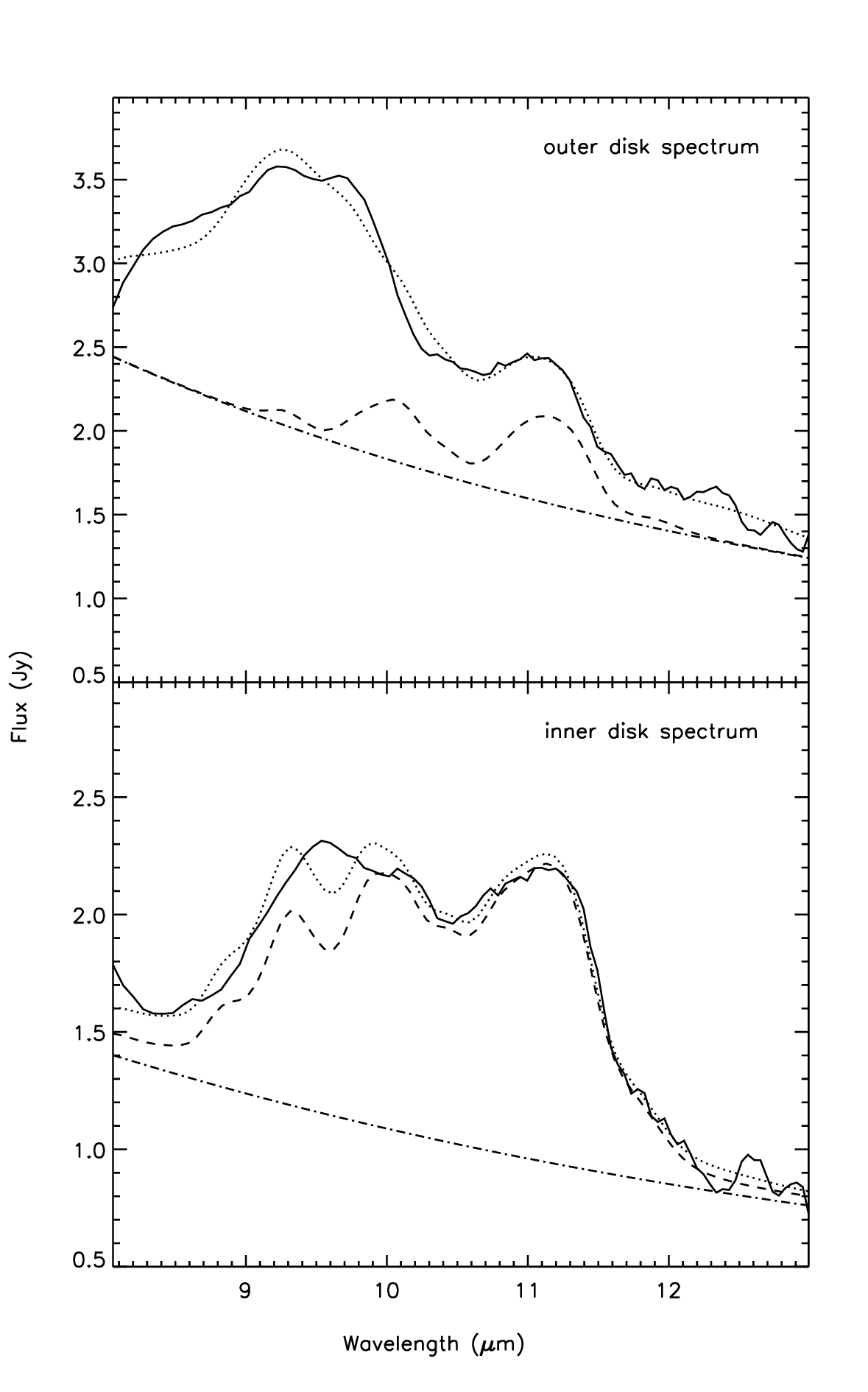}
\caption{The N-band spectrum of the outer disc (top panel) and the
  inner disc (bottom panel) of \object{HD\,52961}, as obtained using a
  projected baseline of 40\,m, are shown using solid lines. The best
  fit models of these spectra are overplotted using a dotted
  line. The continuum and the emission produced by crystalline
  particles are represented using a dashed-dotted and a dashed line
  respectively. For these models, the detailed information is given
  in Table \ref{tab:composition}. It is clear that the hot inner part
  of the disc is much more crystalline than the outer parts.}
\label{fig:chemical_hd52961}
\end{figure}

For \object{HD\,52961}, we fit the spectrum corresponding to the inner disk
(the correlated spectrum, angular size $\sim$ 20 mas) and that
corresponding to the outer disk (the total disk spectrum from which
the correlated spectrum is subtracted) separately. We focus here on
the correlated spectrum taken with the 40\,m baseline. The resulting
best fit model spectra are shown in Fig.~\ref{fig:chemical_hd52961}
and the composition is given in Table~\ref{tab:composition}. The
varying spatial resolution over the N-band introduces an extra slope
in the spectra which was not corrected. Therefore, the characteristic
temperatures ($T_c$) derived for the inner and outer disc spectra are
not realistic. The influence of this slope on the determined chemical
fractions is however expected marginal (see
e.g. \cite{Vanboekel_2004nature}). The average composition over the
disk can be derived by taking the mass weighted average of the inner
(56\%) and outer (44\%) disk regions. The overall composition of the
dust in \object{HD\,52961} is for $\sim$50\% crystalline, and contains
$\sim$60\% $1.5\,\mu$m grains. This is considerably less than what we
find in \object{SX\,Cen}.  It is also clear from Table~\ref{tab:composition}
that the crystalline silicates are not uniformly distributed over the
disk. The inner disk has a much higher crystallinity than the outer
disk.

In order to fit the prominent feature around 9.5\,$\mu$m in the total
and the outer disk spectra of \object{HD\,52961}, we have to add large
(1.5\,$\mu$m) silica grains. We have tried several other dust
components in order to explain this spectral feature, but found no
spectral match using any of them.  We have no explanation for the
presence of these amounts of large silica grains, and thus its
detection is debatable. However, its presence is also indicated from
the Spitzer IRS spectrum, which shows a weak feature around 21\,$\mu$m
which is naturally reproduced using large silica grains (not shown).
For a full mineralogy, the broader wavelength range sampled by our
Spitzer data is needed which is outside the scope of this paper.


\begin{table*}[!t]
\begin{center}
\linespread{1.3}
\selectfont
\begin{tabular}{cccccccccccccc}
\hline
\hline
Star     & T$_c$ &	Cryst.	 & Large	& \multicolumn{2}{c}{Olivine [\%]} & \multicolumn{2}{c}{Pyroxene [\%]} & \multicolumn{2}{c}{Forsterite [\%]} & \multicolumn{2}{c}{Enstatite [\%]} & \multicolumn{2}{c}{Silica [\%]} \\
         & 10$^2$\,K     &	[\%] 			 & grains [\%]	& Small & Large & Small & Large & Small & Large & Small & Large & Small & Large\\
\hline
\object{SX\,Cen} & $6.8_{-0.3}^{+0.3}$ & $78_{-23}^{+17}$ & $93_{-4}^{+3}  $	& - & $21_{-17}^{+24}$ & - & $1_{-1}^{+6}$ & $7_{-3}^{+3}$ & $25_{-9}^{+9}$ & - & $46_{-15}^{+12}$ & $1_{-1}^{+1}$ & $0_{-0}^{+2}$\\
\object{HD\,52961} (inner)&$12_{-2}^{+2}$ & $79_{-12}^{+10}$ & $63_{-12}^{+7} $	& $6_{-6}^{+13}$ & $4_{-4}^{+16}$ & - & $1_{-1}^{+15}$ & $31_{-3}^{+3}$ & $1_{-1}^{+7}$ & $0_{-0}^{+6}$ & $47_{-11}^{+9}$ & - & $9_{-3}^{+3}$\\
\object{HD\,52961} (outer)&$14_{-1}^{+1}$ & $19_{-3}^{+4}  $ & $59_{-17}^{+18}$	& - & $5_{-5}^{+10}$ & $22_{-17}^{+16}$ & $20_{-16}^{+19}$ & $19_{-3}^{+3}$ & - & - & $0_{-0}^{+4}$ & $0_{-0}^{+1}$ & $34_{-3}^{+4}$\\
\hline
\end{tabular}
\end{center}
\linespread{1}
\caption{The composition and grain sizes of the dust in the
circumstellar environments around \object{SX\,Cen} and \object{HD\,52961} as derived
from our fitting procedure. The olivine and pyroxene grains are
amorphous while the forsterite and enstatite are crystalline. For
\object{HD\,52961} 56\% of the dust mass is in the inner disk region. It
should be noted that, as explained in the text, the temperatures
determined for the inner and outer disc spectra of \object{HD\,52961} are not
realistic.}
\label{tab:composition}
\end{table*}

\subsection{formation history of the disc}
The composition of the dust in the circum-binary disc as a function of
distance from the binary can give important clues to its formation
history. In principle, the disc could have been formed by capturing a
"normal" AGB wind, or through non-conservative mass transfer in an
interacting binary. In the wind scenario, it is not unreasonable to
assume that most dust was in the form of amorphous silicates, since
this is the usual dust composition for O-rich AGB stars with a
moderate to high mass loss rate \citep[see e.g.][]{Sloan_1995,Waters_1996L,
Cami_2002}.  In the interacting binary scenario, the dust may or may
not have formed before the material entered the circum-binary disc,
but in any case the thermal history of that dust would have been very
different from that of the wind scenario: the grains were likely at
high temperatures for a long period of time, increasing the chances of
a substantial crystallisation. Therefore the wind scenario predicts a
predominantly amorphous silicate composition, while the interacting
binary scenario more likely produces (highly) crystalline discs.

Once in the disc, both grain aggregation and crystallisation may
occur. Grain aggregation is a strong function of density and thus
would be most efficient in the inner disc regions. Large grains settle
quickly to the mid-plane thus creating a cold mid-plane population of
grains, which we believe is responsible for the millimeter continuum
emission \citep[see e.g.][]{Deruyter_2006}. The inner disc reaches
temperatures above the glass temperature, forcing the grains to
anneal. Therefore, in both the wind and in the interacting binary
scenarios the innermost disc regions are expected to be strongly
crystalline. The two scenarios predict strongly differing radial
gradients in crystallinity however.

The present-day orbital parameters of the binary systems with
circum-binary discs strongly suggest that interaction took place when
the current post-AGB star was on the AGB. Therefore it seems difficult
to imagine that a standard stellar wind formed the discs, and one
would expect the discs on average to have much higher crystallinity
than typical AGB winds. The recent spectral survey by
\cite{Deruyter_2006b} indeed suggests that the circumbinary discs are
much more crystalline than typical AGB outflows. One complicating
factor is that at present not much is known about the composition of
the dust in AGB outflows in the dust forming layers: by far most data
are spatially unresolved and present the final outcome of the dust
formation process in O-rich AGB outflows.

Our MIDI observations indeed confirm that the inner disc region of
\object{HD\,52961} is extremely crystalline, and that the outer disc regions
are less so. At first glance this would suggest the wind scenario is
more likely, but the orbital parameters indicate substantial AGB
interaction. These first MIDI observations thus raise interesting
questions: is the outer disc of \object{HD\,52961} really amorphous and what
kind of disc formation scenario could lead to amorphous grains? Does
this hold for all systems, or is there an orbital separation
dependence? Clearly more study is required to answer these questions.

\subsection{comparison with Herbig Ae/Be stars}\label{sec:comparison}
In \cite{Deruyter_2006}, it is argued that the broad-band SED
characteristics of the discs around binary post-AGB objects are very
similar to the those of the Herbig Ae/Be group II sources. Herbig
Ae/Be stars are intermediate mass pre-main sequence stars surrounded
by remnant material of the star formation process. For these objects,
the existence of a passive circumstellar disc is firmly established
\citep[e.g.][ and references therein]{Waters_1998, Eisner_2003}. The
Herbig Ae/Be stars are subdivided in two groups \citep{Meeus_2001}
with the group I sources showing a rising mid-IR flux excess,
while the group II sources only show a modest mid-IR excess. The
difference in SED characteristics between both groups is attributed to
disc geometry. The mid-IR excess of group I sources is indicative of
the flaring of the outer disc, while the inner rim of the group II
sources shadows the whole disc and no flaring occurs
\citep{Chiang_1997, Dullemond_2001}. \cite{Vanboekel_2004} used the
most recent models \citep{Dullemond_2004} of the discs around Herbig
Ae/Be stars to compute the visibilities to be expected in the MIDI
wavelength range. \cite{Leinert_2004} on the other hand made
observations with the MIDI instrument of several of these objects. We
make a comparison of the results obtained in these publications for
the group II sources and our observations under the assumption of the
similarity of both source geometries.

%

Concerning the continuum radiation, the modelling performed in
\cite{Vanboekel_2004} shows that the size of the disc increases more
rapidly from 8 to 13 $\mu$m than the interferometric resolution
decreases. This provides a visibility curve which is decreasing from 8
to 13 $\mu$m. The observations indeed show this qualitative behaviour,
however some objects, e.g. HD144432, show a rather horizontal
slope. This very similar behaviour is observed for \object{HD\,52961} as
well. The slope of the continuum visibility does not increase as
expected for a uniform disc, it is instead rather constant with
wavelength. 


Concerning the visibility in the feature, the modelling performed by
\cite{Vanboekel_2004} suggests a lowered visibility for the silicate
feature than for the continuum. The disc is irradiated by the central
object and therefore the disc surface is hotter than the disc
midplane. Because the opacity in the silicate resonance band is higher
than in the continuum, one looks less deep into the disc in the
resonance. This results in the fact that in the 10 $\mu$m region, a
larger region in the resonance is seen than in the continuum. The
observations of Herbig Ae/Be stars show a similar qualitative
behaviour, however the visibility decrease is less pronounced. In
fact, \cite{Vanboekel_2004nature} finds that for three Herbig Ae stars
of the sample of \cite{Leinert_2004}, there is a large radial gradient
in the processing. The innermost region of the proto-planetary discs
has a substantially higher crystallinity degree with a shape very
similar to that of comets in our solar system, while the outer region
is clearly less processed. Clearly, the homogenous distribution of
dust adopted in the modelling of these discs is a very crude
approximation. For \object{HD\,52961}, no visibility decrease is observed in
the feature, instead an increase is seen. The spatial distribution of
the dust responsible for the resonance is not homogeneously
distributed, instead the hot inner region of the dust is much more
crystalline than the outer parts
(Sect. \ref{sec:chemical_composition}). This qualitative similarity in
the distribution of the chemical species in the discs around some
Herbig Ae stars and \object{HD\,52961} is surprising in the context of the
completely different formation history of both.

%
%
\section{Conclusions}\label{sect:conclusions}

The main conclusion of our presented MIDI observations is that they
prove the very compact nature of the circumstellar environments of
\object{HD\,52961} and \object{SX\,Cen}. \object{SX\,Cen} is not resolved using a 45\,m baseline,
which gives an upperlimit of only 18\,AU at the estimated
distance. For the well resolved \object{HD\,52961}, the angular size in the
N-band varies between 35 and 55\,mas in a uniform disc approximation,
which translates to a size of 50 and 80\,AU.

Both stars have an effective temperature in the 6000\,K range and
since there is no evidence for a current dusty mass-loss we 
interpret these results as a very stringent proof of the
existence of a stable reservoir near the star. A Keplerian disc seems
the only plausible solution. The dust sublimation temperature is
reached much further out than the binary orbits, hence the discs must
be circumbinary. This is corroborated by the measured size of the
dust-emission region around \object{HD\,52961}.

Given the size of the orbits, the discs were probably formed in a
poorly understood phase of strong binary interaction, when the star
was at giant dimensions. Both discs are O-rich and there is no
evidence for a C-rich component. They were consequently formed prior
to the late AGB evolution where the stars could have changed into
C-stars. The mass of the companion of \object{SX\,Cen} (1.4 -- 1.9 M$_{\odot}$)
is probably within the range of C-star progenitors. We conclude that
the normal single star AGB evolution was shortcut by the presence of a
binary companion. Clearly the formation of a stable Keplerian disc is
a key ingredient in the late evolution of both binaries.

\object{SX\,Cen} is an RV\,Tauri star of photometric class b which shows a long
term variability of the mean magnitude with a period similar to its
orbital period probably due to variable circumstellar extinction in
the line of sight during orbital motion. The inclination cannot be
very small. Additional interferometric data on different projected
angles will be necessary to probe the expected asymmetries. 

The characteristics of the dust grains seem to be very different from
normal single star outflows. This is shown in the mineralogy of the
silicate resonance feature which shows for both objects a highly
crystalline component and a size distribution with a much stronger
component of large ($> 1$\,$\mu$m) grains than what is observed in
outflows of AGB stars. It is not clear whether this reflects the
formation history of the disc or this is due to the longer
processing time of the dust in the Keplerian discs. Our analysis of
\object{HD\,52961} shows that the crystallinity is clearly concentrated in the
hotter inner region of the disc. Crystallisation by annealing is very
temperature dependent and a similar picture arises as what is seen in
the discs around some young stellar objects: the grains in the hot
inner region were subject to a much stronger processing while in the
outer region remained less processed. MIDI as spectrally dispersed
N-band interferometer is an ideal instrument to study the
chemo-physical structure of the inner regions of these discs.

\begin{acknowledgements} 

The authors would like to thank Jeroen Bouwman for the reduction of
the SPITZER/IRS spectrum of \object{HD\,52961} and Bram Acke for the reduction
of the ISO/SWS spectrum of \object{SX\,Cen}. We also like to thank the referee,
K. Ohnaka, for the many valuable comments. We thank the staff of the Geneva
Observatory and the staff of the Instituut voor Sterrenkunde of the
K.U.Leuven for the generous award of time on the Swiss Euler telescope
at La Silla and the Flemish Mercator telescope at La Roque de los
Muchachos respectively. We also thank our colleagues from the
Instituut voor Sterrenkunde for their contribution to the gathering of
the data.  P.D.~and H.V.W.~acknowledge financial support from the Fund
for Scientific Research of Flanders (FWO).
\end{acknowledgements}

\bibliographystyle{aa}

\begin{thebibliography}{56}
\expandafter\ifx\csname natexlab\endcsname\relax\def\natexlab#1{#1}\fi

\bibitem[{{Alcock} {et~al.}(1998){Alcock}, {Allsman}, {Alves}, {Axelrod},
  {Becker}, {Bennett}, {Cook}, {Freeman}, {Griest}, {Lawson}, {Lehner},
  {Marshall}, {Minniti}, {Peterson}, {Pollard}, {Pratt}, {Quinn}, {Rodgers},
  {Sutherland}, {Tomaney}, \& {Welch}}]{Alcock_1998}
{Alcock}, C., {Allsman}, R.~A., {Alves}, D.~R., {et~al.} 1998, \aj, 115, 1921

\bibitem[{{Balick} \& {Frank}(2002)}]{Balick_2002}
{Balick}, B. \& {Frank}, A. 2002, \araa, 40, 439

\bibitem[{{Bogaert}(1994)}]{Bogaert_1994}
{Bogaert}, E. 1994, Ph.D.~Thesis

\bibitem[{{Bouwman} {et~al.}(2001){Bouwman}, {Meeus}, {de Koter}, {Hony},
  {Dominik}, \& {Waters}}]{Bouwman_2001}
{Bouwman}, J., {Meeus}, G., {de Koter}, A., {et~al.} 2001, \aap, 375, 950

\bibitem[{{Bujarrabal} {et~al.}(2005){Bujarrabal}, {Castro-Carrizo}, {Alcolea},
  \& {Nerii}}]{Bujarrabal_2005}
{Bujarrabal}, V., {Castro-Carrizo}, A., {Alcolea}, J., \& {Nerii}, R. 2005,
  \aap, submitted

\bibitem[{{Bujarrabal} {et~al.}(2003){Bujarrabal}, {Neri}, {Alcolea}, \&
  {Kahane}}]{Bujarrabal_2003}
{Bujarrabal}, V., {Neri}, R., {Alcolea}, J., \& {Kahane}, C. 2003, \aap, 409,
  573

\bibitem[{{Cami}(2002)}]{Cami_2002}
{Cami}, J. 2002, Ph.D.~Thesis

\bibitem[{{Chiang} \& {Goldreich}(1997)}]{Chiang_1997}
{Chiang}, E.~I. \& {Goldreich}, P. 1997, \apj, 490, 368

\bibitem[{{Cohen} {et~al.}(2004){Cohen}, {Van Winckel}, {Bond}, \&
  {Gull}}]{Cohen_2004}
{Cohen}, M., {Van Winckel}, H., {Bond}, H.~E., \& {Gull}, T.~R. 2004, \aj, 127,
  2362

\bibitem[{{De Ruyter} {et~al.}(2005{\natexlab{a}}){De Ruyter}, {Van Winckel},
  \& {Dejonghe}}]{Deruyter_2006b}
{De Ruyter}, S., {Van Winckel}, H., \& {Dejonghe}, H. 2005{\natexlab{a}}, \aap,
  in prep.

\bibitem[{{De Ruyter} {et~al.}(2005{\natexlab{b}}){De Ruyter}, {Van Winckel},
  {Dominik}, {Waters}, \& {Dejonghe}}]{Deruyter_2005}
{De Ruyter}, S., {Van Winckel}, H., {Dominik}, C., {Waters}, L.~B.~F.~M., \&
  {Dejonghe}, H. 2005{\natexlab{b}}, \aap, 435, 161

\bibitem[{{De Ruyter} {et~al.}(2005{\natexlab{c}}){De Ruyter}, {Van Winckel},
  {Maas}, {Lloyd Evans}, \& {Dejonghe}}]{Deruyter_2006}
{De Ruyter}, S., {Van Winckel}, H., {Maas}, T., {Lloyd Evans}, T., \&
  {Dejonghe}, H. 2005{\natexlab{c}}, \aap, submitted

\bibitem[{{Dominik} {et~al.}(2003){Dominik}, {Dullemond}, {Cami}, \& {van
  Winckel}}]{Dominik_2003}
{Dominik}, C., {Dullemond}, C.~P., {Cami}, J., \& {van Winckel}, H. 2003, \aap,
  397, 595

\bibitem[{Dorschner {et~al.}(1995)Dorschner, Begemann, Henning, J\"ager, \&
  Mutschke}]{Dorschner1995}
Dorschner, J., Begemann, B., Henning, T., J\"ager, C., \& Mutschke, H. 1995,
  A\&A, 300, 503

\bibitem[{{Dullemond} \& {Dominik}(2004)}]{Dullemond_2004}
{Dullemond}, C.~P. \& {Dominik}, C. 2004, \aap, 417, 159

\bibitem[{{Dullemond} {et~al.}(2001){Dullemond}, {Dominik}, \&
  {Natta}}]{Dullemond_2001}
{Dullemond}, C.~P., {Dominik}, C., \& {Natta}, A. 2001, \apj, 560, 957

\bibitem[{{Eisner} {et~al.}(2003){Eisner}, {Lane}, {Akeson}, {Hillenbrand}, \&
  {Sargent}}]{Eisner_2003}
{Eisner}, J.~A., {Lane}, B.~F., {Akeson}, R.~L., {Hillenbrand}, L.~A., \&
  {Sargent}, A.~I. 2003, \apj, 588, 360

\bibitem[{{Goldsmith} {et~al.}(1987){Goldsmith}, {Evans}, {Albinson}, \&
  {Bode}}]{Goldsmith_1987}
{Goldsmith}, M.~J., {Evans}, A., {Albinson}, J.~S., \& {Bode}, M.~F. 1987,
  \mnras, 227, 143

\bibitem[{{Gustafsson} {et~al.}(1975){Gustafsson}, {Bell}, {Eriksson}, \&
  {Nordlund}}]{Gustafsson_1975}
{Gustafsson}, B., {Bell}, R.~A., {Eriksson}, K., \& {Nordlund}, A. 1975, \aap,
  42, 407

\bibitem[{{Holland} {et~al.}(1999){Holland}, {Robson}, {Gear}, {Cunningham},
  {Lightfoot}, {Jenness}, {Ivison}, {Stevens}, {Ade}, {Griffin}, {Duncan},
  {Murphy}, \& {Naylor}}]{Holland_1999}
{Holland}, W.~S., {Robson}, E.~I., {Gear}, W.~K., {et~al.} 1999, \mnras, 303,
  659

\bibitem[{{Honda} {et~al.}(2004){Honda}, {Kataza}, {Okamoto}, {Miyata},
  {Yamashita}, {Sako}, {Fujiyoshi}, {Ito}, {Okada}, {Sakon}, \&
  {Onaka}}]{Honda_2004}
{Honda}, M., {Kataza}, H., {Okamoto}, Y.~K., {et~al.} 2004, \apjl, 610, L49

\bibitem[{{Honda} {et~al.}(2003){Honda}, {Kataza}, {Okamoto}, {Miyata},
  {Yamashita}, {Sako}, {Takubo}, \& {Onaka}}]{Honda_2003}
---. 2003, \apjl, 585, L59

\bibitem[{{Jaffe}(2004)}]{Jaffe_2004}
{Jaffe}, W.~J. 2004, in New Frontiers in Stellar Interferometry, Proceedings of
  SPIE Volume 5491. Edited by Wesley A. Traub. Bellingham, WA: The
  International Society for Optical Engineering, 2004., p.715

\bibitem[{{J\"ager} {et~al.}(1998){J\"ager}, {Molster}, {Dorschner}, {Henning},
  {Mutschke}, \& {Waters}}]{Jaeger1998}
{J\"ager}, C., {Molster}, F.~J., {Dorschner}, J., {et~al.} 1998, \aap, 339, 904

\bibitem[{{Kholopov} {et~al.}(1999){Kholopov}, {Samus}, {Frolov}, {Goranskij},
  {Gorynya}, {Karitskaya}, {Kazarovets}, {Kireeva}, {Kukarkina}, {Kurochkin},
  {Medvedeva}, {Pastukhova}, {Perova}, {Rastorguev}, \&
  {Shugarov}}]{Kholopov_1999}
{Kholopov}, P.~N., {Samus}, N.~N., {Frolov}, M.~S., {et~al.} 1999, VizieR
  Online Data Catalog, 2214, 0

\bibitem[{{Leinert} {et~al.}(2003){Leinert}, {Graser}, {Przygodda}, {Waters},
  {Perrin}, {Jaffe}, {Lopez}, {Bakker}, {B{\" o}hm}, {Chesneau}, {Cotton},
  {Damstra}, {de Jong}, {Glazenborg-Kluttig}, {Grimm}, {Hanenburg}, {Laun},
  {Lenzen}, {Ligori}, {Mathar}, {Meisner}, {Morel}, {Morr}, {Neumann}, {Pel},
  {Schuller}, {Rohloff}, {Stecklum}, {Storz}, {von der L{\" u}he}, \&
  {Wagner}}]{Leinert_2003}
{Leinert}, C., {Graser}, U., {Przygodda}, F., {et~al.} 2003, \apss, 286, 73

\bibitem[{{Leinert} {et~al.}(2004){Leinert}, {van Boekel}, {Waters},
  {Chesneau}, {Malbet}, {K{\" o}hler}, {Jaffe}, {Ratzka}, {Dutrey},
  {Preibisch}, {Graser}, {Bakker}, {Chagnon}, {Cotton}, {Dominik}, {Dullemond},
  {Glazenborg-Kluttig}, {Glindemann}, {Henning}, {Hofmann}, {de Jong},
  {Lenzen}, {Ligori}, {Lopez}, {Meisner}, {Morel}, {Paresce}, {Pel},
  {Percheron}, {Perrin}, {Przygodda}, {Richichi}, {Sch{\" o}ller}, {Schuller},
  {Stecklum}, {van den Ancker}, {von der L{\" u}he}, \&
  {Weigelt}}]{Leinert_2004}
{Leinert}, C., {van Boekel}, R., {Waters}, L.~B.~F.~M., {et~al.} 2004, \aap,
  423, 537

\bibitem[{{Lucy} \& {Sweeney}(1971)}]{Lucy_1971}
{Lucy}, L.~B. \& {Sweeney}, M.~A. 1971, \aj, 76, 544

\bibitem[{{Maas} {et~al.}(2003){Maas}, {Van Winckel}, {Lloyd Evans}, {Nyman},
  {Kilkenny}, {Martinez}, {Marang}, \& {van Wyk}}]{Maas_2003}
{Maas}, T., {Van Winckel}, H., {Lloyd Evans}, T., {et~al.} 2003, \aap, 405, 271

\bibitem[{{Maas} {et~al.}(2002){Maas}, {Van Winckel}, \&
  {Waelkens}}]{Maas_2002}
{Maas}, T., {Van Winckel}, H., \& {Waelkens}, C. 2002, \aap, 386, 504

\bibitem[{{Meeus} {et~al.}(2001){Meeus}, {Waters}, {Bouwman}, {van den Ancker},
  {Waelkens}, \& {Malfait}}]{Meeus_2001}
{Meeus}, G., {Waters}, L.~B.~F.~M., {Bouwman}, J., {et~al.} 2001, \aap, 365,
  476

\bibitem[{{Men'shchikov} {et~al.}(2002){Men'shchikov}, {Schertl}, {Tuthill},
  {Weigelt}, \& {Yungelson}}]{Menshchikov_2002}
{Men'shchikov}, A.~B., {Schertl}, D., {Tuthill}, P.~G., {Weigelt}, G., \&
  {Yungelson}, L.~R. 2002, \aap, 393, 867

\bibitem[{{Min} {et~al.}(2003){Min}, {Hovenier}, \& {de Koter}}]{Min_2003}
{Min}, M., {Hovenier}, J.~W., \& {de Koter}, A. 2003, \aap, 404, 35

\bibitem[{{Min} {et~al.}(2005){Min}, {Hovenier}, \& {de Koter}}]{Min_2005}
---. 2005, \aap, 432, 909

\bibitem[{{O'Connell}(1933)}]{Oconnell_1933}
{O'Connell}, D.~J.~K. 1933, Harvard Bull., 893, 14

\bibitem[{{Savage} \& {Mathis}(1979)}]{Savage_1979}
{Savage}, B.~D. \& {Mathis}, J.~S. 1979, \araa, 17, 73

\bibitem[{Servoin \& Piriou(1973)}]{Servoin1973}
Servoin, J.~L. \& Piriou, B. 1973, Phys. Stat. Sol. (b), 55, 677

\bibitem[{{Shenton} {et~al.}(1994){Shenton}, {Evans}, {Albinson}, {Barrett},
  {Davies}, {Goldsmith}, {Hutchinson}, {Laney}, \& {Maddison}}]{Shenton_1994}
{Shenton}, M., {Evans}, A., {Albinson}, J.~S., {et~al.} 1994, \aap, 292, 102

\bibitem[{{Sloan} \& {Price}(1995)}]{Sloan_1995}
{Sloan}, G.~C. \& {Price}, S.~D. 1995, \apj, 451, 758

\bibitem[{Spitzer \& Kleinman(1960)}]{1960PhRv..121.1324S}
Spitzer, W.~G. \& Kleinman, D.~A. 1960, Physical Review, 121, 1324

\bibitem[{{Tubbs} {et~al.}(2004){Tubbs}, {Meisner}, {Bakker}, \&
  {Albrecht}}]{Tubbs_2004}
{Tubbs}, R.~N., {Meisner}, J.~A., {Bakker}, E.~J., \& {Albrecht}, S. 2004, in
  New Frontiers in Stellar Interferometry, Proceedings of SPIE Volume 5491.
  Edited by Wesley A. Traub. Bellingham, WA: The International Society for
  Optical Engineering, 2004., p.588

\bibitem[{{van Boekel}(2004)}]{Vanboekel_2004}
{van Boekel}, R. 2004, PhD thesis, University of Amsterdam

\bibitem[{{van Boekel} {et~al.}(2004){van Boekel}, {Min}, {Leinert}, {Waters},
  {Richichi}, {Chesneau}, {Dominik}, {Jaffe}, {Dutrey}, {Graser}, {Henning},
  {de Jong}, {K{\" o}hler}, {de Koter}, {Lopez}, {Malbet}, {Morel}, {Paresce},
  {Perrin}, {Preibisch}, {Przygodda}, {Sch{\" o}ller}, \&
  {Wittkowski}}]{Vanboekel_2004nature}
{van Boekel}, R., {Min}, M., {Leinert}, C., {et~al.} 2004, \nat, 432, 479

\bibitem[{{van Boekel} {et~al.}(2005){van Boekel}, {Min}, {Waters}, {de Koter},
  {Dominik}, {van den Ancker}, \& {Bouwman}}]{Vanboekel_2005}
{van Boekel}, R., {Min}, M., {Waters}, L.~B.~F.~M., {et~al.} 2005, \aap, 437,
  189

\bibitem[{{Van Winckel}(2003)}]{Vanwinckel_2003}
{Van Winckel}, H. 2003, \araa, 41, 391

\bibitem[{{Van Winckel} {et~al.}(1992){Van Winckel}, {Mathis}, \&
  {Waelkens}}]{Vanwinckel_1992}
{Van Winckel}, H., {Mathis}, J.~S., \& {Waelkens}, C. 1992, \nat, 356, 500

\bibitem[{{Van Winckel} {et~al.}(1999){Van Winckel}, {Waelkens}, {Fernie}, \&
  {Waters}}]{Vanwinckel_1999}
{Van Winckel}, H., {Waelkens}, C., {Fernie}, J.~D., \& {Waters}, L.~B.~F.~M.
  1999, \aap, 343, 202

\bibitem[{{Van Winckel} {et~al.}(1995){Van Winckel}, {Waelkens}, \&
  {Waters}}]{Vanwinckel_1995}
{Van Winckel}, H., {Waelkens}, C., \& {Waters}, L.~B.~F.~M. 1995, \aap, 293,
  L25

\bibitem[{{Verhoelst}(2005)}]{Verhoelst_2005}
{Verhoelst}, T. 2005, Ph.D.~Thesis

\bibitem[{{Vo\^ute}(1940)}]{Voute_1940}
{Vo\^ute}, J. 1940, Lembang Ann., 8, 42

\bibitem[{{Waelkens} {et~al.}(1991{\natexlab{a}}){Waelkens}, {Lamers},
  {Waters}, {Rufener}, {Trams}, {Le Bertre}, {Ferlet}, \&
  {Vidal-Madjar}}]{Waelkens_1991a}
{Waelkens}, C., {Lamers}, H.~J.~G.~L.~M., {Waters}, L.~B.~F.~M., {et~al.}
  1991{\natexlab{a}}, \aap, 242, 433

\bibitem[{{Waelkens} {et~al.}(1991{\natexlab{b}}){Waelkens}, {Van Winckel},
  {Bogaert}, \& {Trams}}]{Waelkens_1991}
{Waelkens}, C., {Van Winckel}, H., {Bogaert}, E., \& {Trams}, N.~R.
  1991{\natexlab{b}}, \aap, 251, 495

\bibitem[{{Waters} {et~al.}(1996){Waters}, {Molster}, {de Jong}, {Beintema},
  {Waelkens}, {Boogert}, {Boxhoorn}, {de Graauw}, {Drapatz}, {Feuchtgruber},
  {Genzel}, {Helmich}, {Heras}, {Huygen}, {Izumiura}, {Justtanont}, {Kester},
  {Kunze}, {Lahuis}, {Lamers}, {Leech}, {Loup}, {Lutz}, {Morris}, {Price},
  {Roelfsema}, {Salama}, {Schaeidt}, {Tielens}, {Trams}, {Valentijn},
  {Vandenbussche}, {van den Ancker}, {van Dishoeck}, {Van Winckel},
  {Wesselius}, \& {Young}}]{Waters_1996L}
{Waters}, L.~B.~F.~M., {Molster}, F.~J., {de Jong}, T., {et~al.} 1996, \aap,
  315, L361

\bibitem[{{Waters} {et~al.}(1992){Waters}, {Trams}, \&
  {Waelkens}}]{Waters_1992}
{Waters}, L.~B.~F.~M., {Trams}, N.~R., \& {Waelkens}, C. 1992, \aap, 262, L37

\bibitem[{{Waters} \& {Waelkens}(1998)}]{Waters_1998}
{Waters}, L.~B.~F.~M. \& {Waelkens}, C. 1998, \araa, 36, 233

\bibitem[{{Waters} {et~al.}(1993){Waters}, {Waelkens}, {Mayor}, \&
  {Trams}}]{Waters_1993}
{Waters}, L.~B.~F.~M., {Waelkens}, C., {Mayor}, M., \& {Trams}, N.~R. 1993,
  \aap, 269, 242

\end{thebibliography}

\end{document}